\documentclass[12pt]{iopart}
\usepackage{graphicx}
\usepackage{iopams}
\graphicspath{{},{figures/}}
\begin{document}

\title[Hour-glass dispersion in Fe$_{1+y}$Te$_{0.7}$Se$_{0.3}$ and its relation to high-$T_c$ superconductivity]{Magnetic hour-glass dispersion and its relation to high-temperature superconductivity in iron-tuned Fe$_{1+y}$Te$_{0.7}$Se$_{0.3}$}

\author{N. Tsyrulin$^1$, R. Viennois$^{2,3}$, E. Giannini$^3$, M. Boehm$^4$, M. Jimenez-Ruiz$^4$, A. A. Omrani$^1$, B. Dalla Piazza$^1$ and H. M. R\o{}nnow$^1$ }
\address{$^1$ Laboratory for Quantum Magnetism,  \'{E}cole Polytechnique F\'{e}d\'{e}rale de Lausanne (EPFL), CH-1015 Lausanne, Switzerland}
\address{$^2$ Institut Charles Gerhardt Montpellier, Universit\'{e} Montpellier II, F-34095 Montpellier, France}
\address{$^3$ DPMC, University of Geneva, 24 Quai E.-Ansermet, 1211 Geneva, Switzerland}
\address{$^4$ Institut Laue Langevin, 6 rue Jules Horowitz BP156, 38024 Grenoble CEDEX 9, France}
\begin{abstract}

High-temperature superconductivity remains arguably the largest outstanding enigma of condensed matter physics.
The discovery of iron-based high-temperature superconductors \cite{Kamihara2008,Ren} has renewed the importance of understanding superconductivity in materials susceptible to magnetic order and fluctuations.
Intriguingly they show magnetic fluctuations reminiscent of the superconducting (SC) cuprates \cite{tranquada_review}, including a 'resonance' and an 'hour-glass' shaped dispersion~\cite{Li}, which provide an opportunity to new insight to the coupling between spin fluctuations and superconductivity. Here we report inelastic neutron scattering data on Fe$_{1+y}$Te$_{0.7}$Se$_{0.3}$ using excess iron concentration to tune between a SC ($y=0.02$) and a non-SC ($y=0.05$) ground states.
We find incommensurate spectra in both samples but discover that in the one that becomes SC, a constriction towards a commensurate hourglass shape develop well above $T_c$. Conversely a spin-gap and concomitant spectral weight shift happen below $T_c$. Our results imply that the hourglass shaped dispersion is most likely a pre-requisite for superconductivity, whereas the spin-gap and shift of spectral weight are consequences of superconductivity. We explain this observation by pointing out that an inwards dispersion towards the commensurate wave-vector is needed for the opening of a spin gap to lower the magnetic exchange energy and hence provide the necessary condensation energy for the SC state to emerge.
\end{abstract}
\maketitle

Essentially most families of SC cuprates display a common spin excitation spectrum with a spin resonance, an hour-glass shape dispersion and a spin gap below $T_c$.
This magnetic hour-glass dispersion is quite unusual and seems almost concomitant to high-temperature superconductivity, although few non-SC examples were recently reported \cite{Ulbrich}. However, despite intense debate based on extensive theoretical and experimental efforts, the key question has remained elusive: what is the relation between the hour-glass and the SC transition - and which is a consequence of the other?
Surprisingly the recently discovered iron-based superconductors display remarkable similarities in the magnetic excitation spectrum strongly suggesting a common mechanism of superconductivity. In iron-arsenic based compounds as well as in the iron chalcogenides emergence of superconductivity is accompanied by opening of a spin gap and appearance of a commensurate 'spin resonance' and 'hour-glass' shaped magnetic dispersion which have been observed in several inelastic neutron scattering experiments \cite{Johnston,Lumsden_rew,Inosov}. We have discovered that tuning superconductivity in iron chalcogenide by small amounts of excess iron provides fresh insight to the mechanism of SC.

Fe$_{1+y}$Te$_{0.7}$Se$_{0.3}$ displays the simplest single layered crystal structure among the iron-based superconductors. Excess of iron located on an interstitial site strongly influences both magnetic and SC properties as a tuning parameter in addition to the Te/Se ratio \cite{Liu,Viennois1,Bendele}.
For our experiments, two batches of Fe$_{1+y}$Te$_{0.7}$Se$_{0.3}$ were grown by Bridgman-Stockbarger method with starting $Fe:(Te,Se)$ ratios of $1:1$ and $0.9:1$ respectively. Precursors were heated up to $T=930^\circ\;\mathrm{C}$ in quartz tubes and then slowly cooled. Structure refinement of small single-crystals determined that the iron contents $x$ is $y=0.05$ and $y=0.02$ for the two batches, respectively. This difference of iron content was confirmed by EDX analysis. More details can be found in ref. \cite{Viennois1}.

\begin{figure}
\begin{center}
\includegraphics*[width=0.75\columnwidth,bbllx=0,bblly=0,bburx=1,bbury=1.2,angle=0,clip=]{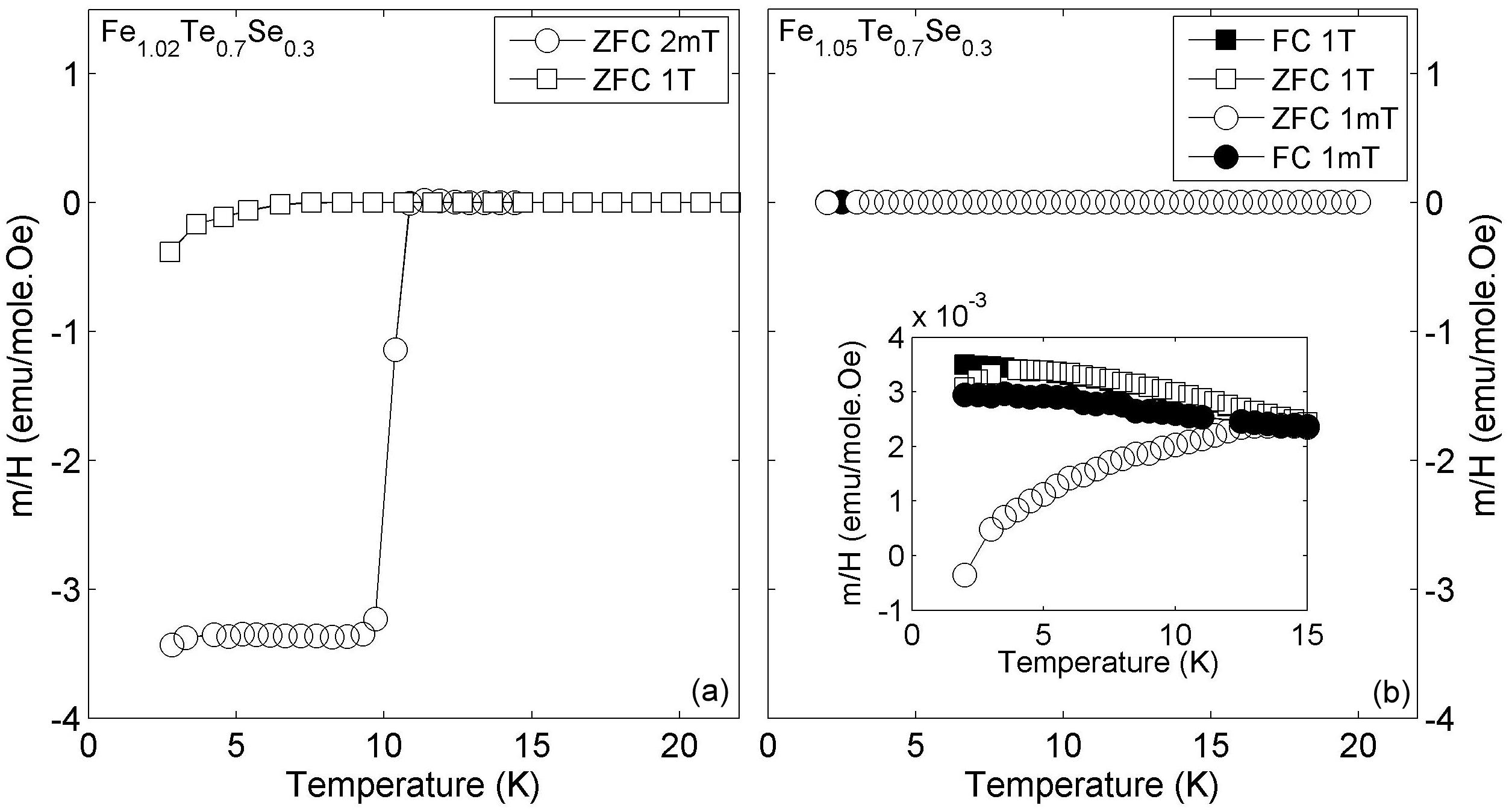}
  \caption{Magnetization of Fe$_{1.02}$Te$_{0.7}$Se$_{0.3}$ and Fe$_{1.05}$Te$_{0.7}$Se$_{0.3}$ measured as function of temperature is shown in (a) and (b), respectively. White and black circles show the result of low-field ZFC and FC measurements, respectively. White and black squares represent the results of ZFC and FC measurements performed at $\mu_0H=1\;\mathrm{T}$.}
  \label{mag}
  \end{center}
\end{figure}

Superconductivity was determined by the Meissner effect in magnetization measurements on single crystals of Fe$_{1.05}$Te$_{0.7}$Se$_{0.3}$ and Fe$_{1.02}$Te$_{0.7}$Se$_{0.3}$ with masses of 7.4(1)~mg and 8.1(1)~mg, respectively.
Measurements were performed with magnetic field applied along the crystallographic $c$-axis after zero-field cooling (ZFC) or field cooling (FC).
At $\mu_0H=2\;\mathrm{mT}$ a sharp drop of ZFC magnetization in the $y=0.02$ sample demonstrate bulk superconductivity developing between $T_c^{onset}=10.8$~K and $T_c^{bulk}=9.7$~K
(figure\ \ref{mag}(a)). The sharpness of the transition witness high sample homogeneity and quality, which may explain the sharpness of our inelastic data compared to previous studies on larger crystals.
In contrast, the $y=0.05$ sample show essentially no Meissner effect (figure\ \ref{mag}(b)). Comparing to $y=0.02$, the very small decrease in ZFC magnetization of 0.03~emu/(mole Oe) implies that any superconducting volume fraction in our Fe$_{1.05}$Te$_{0.7}$Se$_{0.3}$ sample is smaller than of 0.1\%.
Thus we have successfully prepared two high quality samples with same Se-content and a controlled small Fe-content tuning the system from a homogenous bulk superconductor at $y=0.02$ to a completely non-superconducting sample at $y=0.05$.

The spin dynamics in both samples was investigated using the thermal neutron three axis spectrometer IN8 at the
Institut Laue-Langevin, France.
Three single crystals with iron content $y=0.02$ with a total mass of
$m=0.85(3)\;\mathrm{g}$ were co-aligned with a total mosaic spread less than 0.7$^\circ$. One single crystal $y=0.05$ with a mass $m=0.63(1)\;\mathrm{g}$ was used for the measurements. Both samples were fixed on aluminum holders.
The reciprocal plane $(h, k, 0)$ of both compounds was co-aligned with the horizontal scattering plane of the spectrometer. The measurements were performed using a standard $^4$He Orange cryostat in the temperature range $2\;\mathrm{K}\leq T \leq 40\;\mathrm{K}$.
Using (002) Bragg reflection from a pyrolithic graphite (PG) analyzer, the final energy of neutrons were set to
$E_f=14\;\mathrm{meV}$. We used PG(002) as a monochromator and double focusing mode of the spectrometer.
Such a configuration resulted in the energy and the wave-vector resolution of
$\Delta E \approx1\;\mathrm{meV}$ and $\Delta Q \approx 0.063\;\mathrm{\AA^{-1}}$, respectively which was measured by performing an energy scan through an incoherent position and a wave-vector (Q-) scan across a nuclear Bragg peak, respectively. Measured intensity was normalized to the incident flux monitor.
\par
\begin{figure}[htbp]
\begin{center}
\includegraphics*[width=0.75\columnwidth,bbllx=0,bblly=0,bburx=1,bbury=1.2,angle=0,clip=]{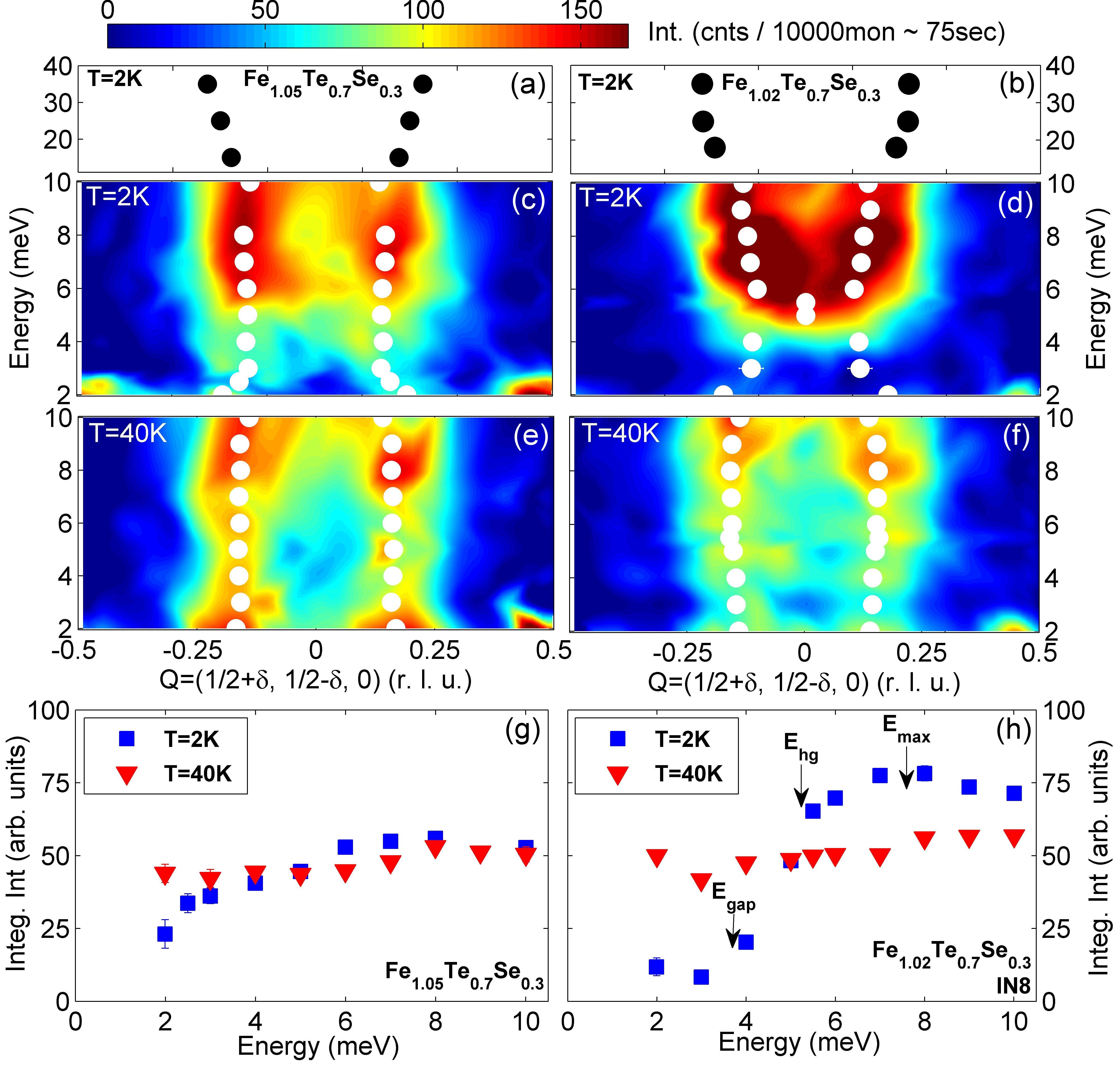}
\caption{Magnetic spectrum and dispersion along $(1/2+\delta, 1/2-\delta, 0)$ in $y=0.05$ (a,c,e) and $y=0.02$ (b,d,f), from $Q$-scans at a series of energies. Colormaps represent intensity with a Q-independent bacground subtracted.
Below 2~meV, incoherent elastic background becomes dominant.
Data above $E=10\;\mathrm{meV}$ were measured in different configuration and intensities are not directly comparable.
Each scan was fitted by two symmetric Gaussians, yielding the dispersion (cirles) and integrated intensity, which is shown in (g,h) as function of energy. Blue and red circles correspond to  $T=2$~K and at $T=40$~K, respectively.}
\label{CP}
\end{center}
\end{figure}

Our inelastic neutron scattering measurements revealed steeply dispersing magnetic excitations at positions $(1/2\pm\delta, 1/2\mp\delta, 0)$ in both samples. The peak widths exceed the instrumental resolution in agreement with observations in related materials \cite{Argyriou}.
The dispersion was mapped by performing constant energy scans as summarized in figure~\ref{CP}. Each scan was fitted by two Gaussians symmetrically  placed around (1/2, 1/2 ,0). The resulting peak-positions are indicated as points on top of the colormaps.

Starting with the non-SC $y=0.05$ sample, at $T=40\;\mathrm{K}$ the incommensuration $\delta=0.154(14)$ and intensity is essentially energy independent up to 10~meV.
This value of $\delta$ and the dispersion above $E=10$~meV is consistent with previous reports \cite{Qiu,Wen,Argyriou,Babkevich}.
Lowering temperature to $T=2\;\mathrm{K}$ has minor influence on the incommensuration, and no commensurate signal was detected in this sample up to $E=35\;\mathrm{meV}$.
At $T=40\;\mathrm{K}$, the $y=0.02$ sample display very similar incommensuration. However, lowering the temperature to $T=2\;\mathrm{K}$ reveals dramatic differences:
i) Q-scans at $E=5$~meV and $E=5.5$~meV narrow into a single commensurate peak defining $E_{hg}=5.3(5)$~meV; ii) spectral weight is removed below $E_{hg}$ and shifted to above $E_{hg}$.
This dramatic restructuring of the magnetic excitation spectrum is a direct experimental evidence of intricate coupling between magnetism and superconductivity,
and is very reminiscent of the behavior in the cuprate superconductors, with one noticeable difference: In Fe$_{1.02}$Te$_{0.7}$Se$_{0.3}$ the spectrum is completely incommensurate at high temperature and becomes commensurate upon lowering temperature. In the cuprates the hourglass-shape was thought to persists at all temperatures, and our discovery calls for an investigation versus temperature \emph{e.g.} in underdoped La$_{2-x}$Sr$_x$CuO$_4$.

\begin{figure}
\begin{center}
  \includegraphics*[width=0.7\columnwidth,bbllx=0,bblly=0,bburx=1,bbury=1.2,angle=0,clip=]{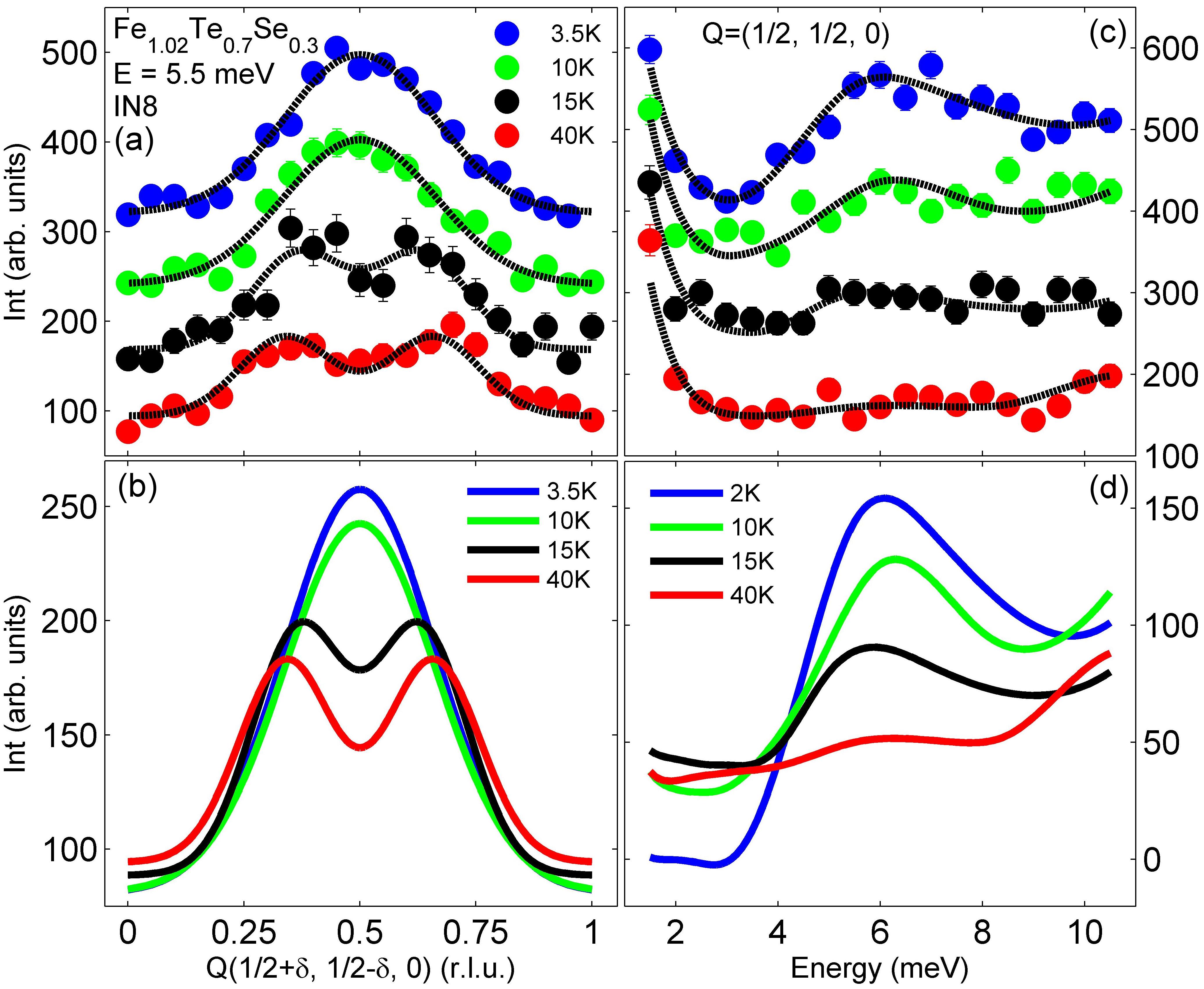}
  \caption{(a) Scans at constant $E=5.5\;\mathrm{meV}$ for $T=3.5\;\mathrm{K}$, $T=10\;\mathrm{K}$, $T=15\;\mathrm{K}$ and
$T=40\;\mathrm{K}$ are shown by red, black, green, brown and blue circles, respectively. For clarity, scans are displaced vertically. Black lines are fits to two Gaussians symmetric around (1/2,1/2,0). (b) The comparison of fits results for each temperature. (c) Constant Q-scans performed at $Q=(1/2, 1/2, 0)$ at $T=2\;\mathrm{K}$, $T=10\;\mathrm{K}$, $T=15\;\mathrm{K}$ and $T=40\;\mathrm{K}$. For clarity, scans are displaced vertically. Lines are smooth curves aiding comparison between temperatures. (d) The smooth curves compared for the different temperatures, with the tail from the elastic line removed.
}
  \label{rawdata}
  \end{center}
\end{figure}

Our data are consistent with those previously reported for SC FeSe$_{0.4}$Te$_{0.6}$ and "nearly superconducting" (NSC) FeSe$_{0.45}$Te$_{0.55}$~\cite{Li}. In SC FeSe$_{0.4}$Te$_{0.6}$ a constriction towards commensuration was also observed at $E=4.5$~meV as temperature is lowered from $T=20$~K to $T=1.5$~K. It is possible that also that sample would show complete incommensurability at $T=40$~K. In ref.~\cite{Li} the NSC data were interpreted to also show an hourglass dispersion, which led to the conclusion that it is not directly associated with superconductivity. However, with the new insight provided by our non-superconducting Fe$_{1.05}$Te$_{0.7}$Se$_{0.3}$ data, the NSC data of ref.~\cite{Li} can be reassessed to be consistent with our interpretation.
Indeed at $T=15$~K, there is no discernible commensuration at $E=4.5$~meV, and there is little – if any – commensuration upon lowering temperature to $T=4$~K. The explanation for why our non-superconducting sample display a sharper clearly incommensurate spectrum may be that it is completely non-superconducting, and much smaller (0.63g compared to 23g in ref.~\cite{Li}), therefore likely more homogenous. The NSC sample in ref.~\cite{Li} had up to 30$\%$ superconducting volume fraction.

\par

Comparing to the cuprates, we argue that three characteristic energies must be defined: $E_{gap}$ below which spectral weight is depleted in the SC state; $E_{max}$ - the energy where a maximum in intensity develops in the SC state; and $E_{hg}$ the energy where the incommensurate spectrum constricts towards the commensurate wave-vector, thereby forming an hour-glass shape. In both La$_{2-x}$Sr$_x$CuO$_4$ and YBCO - the two cuprate families most studied by inelastic neutron scattering - there are incommensurate excitations with an 'hour-glass' shaped dispersion towards the commensurate point at $E_{hg}$ around $40$~meV \cite{Vignolle,Reznik}. In La$_{2-x}$Sr$_x$CuO$_4$ the spin gap $E_{gap}$ that opens upon entering the SC state is only about 8~meV at optimal doping, and spectral weight is redistributed along the incommensurate parts of the excitations, such that a maximum in intensity occurs at an incommensurate resonance around $E_{max}=18\;\mathrm{meV}\sim2E_{gap}$ \cite{christensen2004,tranquada2004}. The maximum in intensity is hence a consequence of shifting weight from below the gap. Whereas $E_{gap}$ and $E_{max}$ are therefore two manifestations of the same feature, in La$_{2-x}$Sr$_x$CuO$_4$ the commensurate energy of the hour-glass dispersion $E_{hg}$ is independent hereof. In YBCO on the other hand, the energy scale of $E_{gap}$ is higher such that the redistribution of spectral weight falls around the commensurate energy such that $E_{max}$ and $E_{hg}$ tend to coincide into a pronounced commensurate "resonance" peak.

\begin{figure}
\begin{center}
  \includegraphics*[width=0.8\columnwidth,bbllx=0,bblly=0,bburx=1,bbury=1.2,angle=0,clip=]{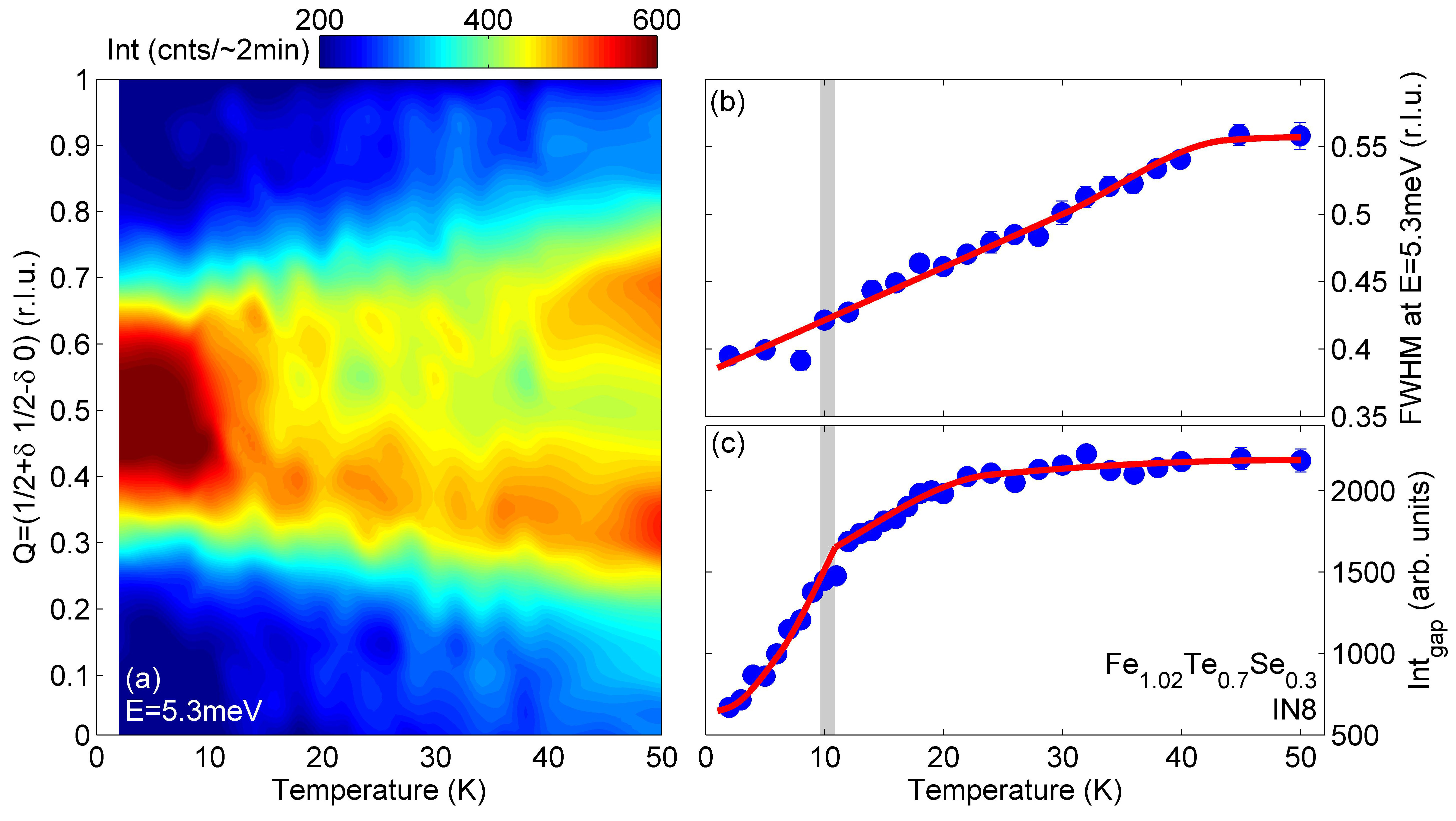}
  \caption{(a) Temperature evolution of constant energy scan at $E=5.3$~meV is shown as a colormap. The figure is obtained by merging and smoothing twenty one Q-scan in the temperature range from $T=2.5$~K to $T=50$~K in total.
  (b) Temperature dependence of the full-width-half-max of $Q$-scans at constant $E=5.3$~meV.The shift towards the commensurate center point happens gradually from 40~K and down. Red line is a guide to an eye. (c) Intensity at 3~meV below $E_{\mathrm{gap}}$ as function of temperature. A pronounced onset of depletion of intensity sets in below $T_c$, consistent with the opening of a gap, whereas there is only weak temperature dependence above $T_c$. Red line shows a guide to an eye.}
    \label{Intdata}
    \end{center}
\end{figure}

To quantify the details of the spectral weight restructuring, we plot in figures~\ref{CP}(g,~h) the integrated magnetic intensity extracted from Gaussian fits to the constant energy scans. For $y=0.05$, intensity is almost energy independent. The slight difference between $T=40\;\mathrm{K}$ and $T=2\;\mathrm{K}$ is consistent with $1-e^{-E/k_BT}$, implying an essentially temperature independent dynamic susceptibility.
For $y=0.02$, the $T=40\;\mathrm{K}$ intensity has similar weak energy dependence, but the $T=2\;\mathrm{K}$ intensity show a sharp spin gap of $E_{gap}=3.7(5)$~meV, with only weak intensity remaining below. It can be seen that the spectral weight removed from low energy is shifted to energies above $E_{hg}$, creating a maximum at $E_{max}=7.5(5)$~meV.
There is less increase in the $Q$-integrated intensity at the commensurate position $E_{hg}=5.3(5)\;\mathrm{meV}$, where the increase in peak intensity at the commensurate position comes from the merging in $Q$ of two incommensurate peaks, and not from energies below or above $E_{hg}$. In this respect the behavior of Fe$_{1.02}$Te$_{0.7}$Se$_{0.3}$ with $E_{hg}\lesssim E_{max}$ is more akin to YBCO with $E_{hg}\simeq E_{max}$ than to La$_{2-x}$Sr$_x$CuO$_4$ with $E_{hg}> E_{max}$.

\begin{figure*}
\includegraphics*[width=\textwidth,bbllx=0,bblly=0,bburx=1,bbury=1.2,angle=0,clip=]{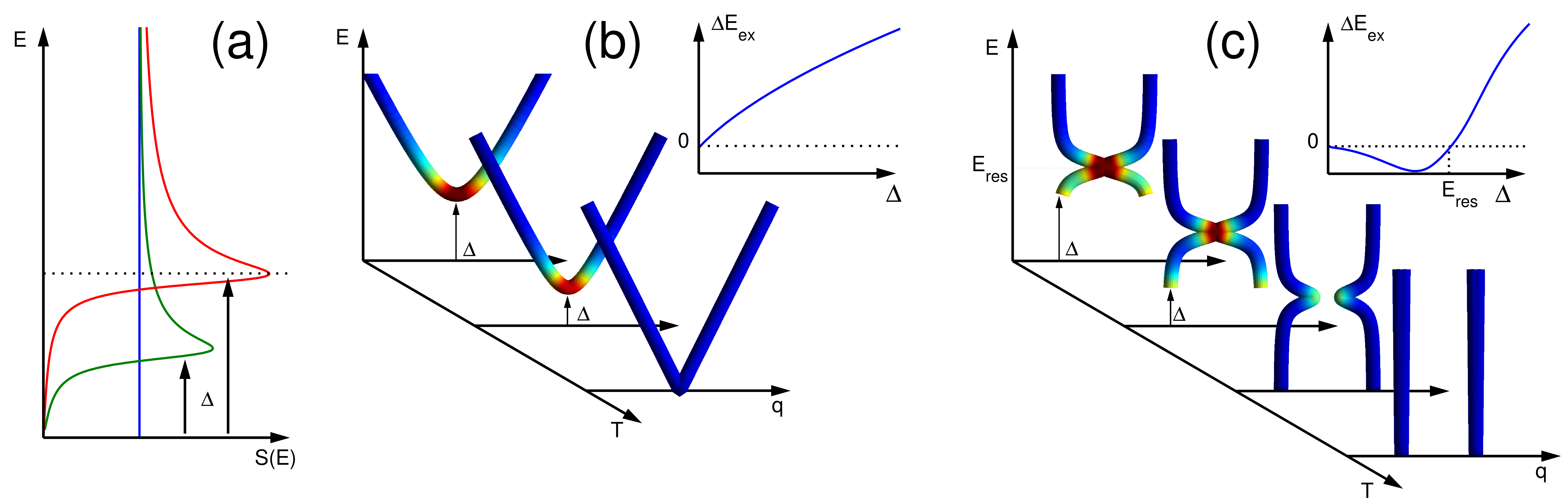}
\caption{Sketch of how opening an energy gap in the spin excitation spectrum changes the magnetic exchange energy $\Delta{}E_{ex}$. (a) The energy dependent scattering function $S(E)$, which for a 2D antiferromagnet is constant in absence of a gap. When the gap opens spectral weight is shifted from below the gap to above the gap. (b) Opening a gap in a conventional linear dispersion shifts spectral weight away from the commensurate $(\pi,\pi)$ point, and would increase $\Delta{}E_{ex}$ (inset). (c) Opening a gap in an hour-glass dispersion shifts spectral weight towards $q=(\pi,\pi)$, and creates a minimum in $\Delta{}E_{ex}$ for $\Delta$ just below $E_{\mathrm{res}}$. }
\label{fig:sketch}
\end{figure*}

The temperature dependence of the spectral restructuring was studied by performing $Q$-scans at constant energy $E=5.5$~meV and energy-scans at $Q=(1/2,1/2,0)$.
Figure~\ref{rawdata}(a) show $Q$-scans fitted  by Gaussian peaks with a linear background. For clarity we plot the fit curves separately  in figure~\ref{rawdata}(b).
Importantly, the peaks already start shifting towards the commensurate position at $T=15\;\mathrm{K}$, well above $T_c$ and the commensuration is almost complete already at $ T=10\;\mathrm{K}\simeq T_{c}$, with little subsequent change down to $T=3.5\;\mathrm{K}$. A color plot in figure~\ref{Intdata}(a) shows a complete evolution of the $Q$-scan performed at $E_{hg}=5.3$~meV in the temperature range from $T=50\;\mathrm{K}$ to $T=2.5\;\mathrm{K}$. This data together with the one shown in figure~\ref{CP}(d,~f) clearly illustrates the process of commensuration towards an hour-glass shape dispersion as function of temperature. The constriction towards commensuration at $E_{hg}=5.3$~meV is clearly develops \emph{above} $T_c$. This is further documented in figure \ref{Intdata}(b), showing how the full-width at half-maximum of the double peak structure at $E_{hg}=5.3$~meV decreases continuously from $T=40$~K to $T=2.5$~K with majority of the commensuration happening before T$_c$.

In contrast, the energy scans displayed in figure~\ref{rawdata}(c-d) show a different temperature dependence for the spin gap $E_{gap}=3.7$~meV.
Below $E=3.7\;\mathrm{meV}$, intensity remains constant from $T=40\;\mathrm{K}$ to $T=10\;\mathrm{K}$, and only \emph{below} $T_c$ does intensity suddenly drop.
This is also seen in a temperature scan at $E=3$~meV (figure~\ref{Intdata}(b)), which show a very different temperature dependence than figure~\ref{Intdata}(a), with a kink to rapid depletion of intensity below $T_c$. The weak decrease above $T_c$ may be accounted for by proximity to the 5.3~meV commensuration, or a fluctuating pre-cursor to the gap.
Hence, the spin gap opens sharply \emph{below} $T_c$. Combined with the fact that superconductivity is absent in Fe$_{1.05}$Te$_{0.7}$Se$_{0.3}$, in which the magnetic spectrum remain incommensurate at all temperatures, the implications of our observations is that the commensurate hour-glass shape at $E_{hg}$ of the magnetic spectrum is most likely a pre-requisite to superconductivity, whereas the spin-gap is a consequence of superconductivity, and the concomitant maximum in intensity at $E_{max}$ is in turn a consequence of the spin-gap.

The observation that the hour glass dispersion seem a necessary condition for superconductivity supports that the SC condensation energy can come from the change in magnetic exchange energy between the normal and SC state \cite{Scalapino98,Woo2006}. Based on arguments from the $t-J$ model, whose essence should be extendable to Fe-based system, the change in exchange energy can be written as $\Delta{}E_{ex}=2J(\langle \mathbf{S}_i\cdot\mathbf{S}_j\rangle_S-\langle \mathbf{S}_i\cdot\mathbf{S}_j\rangle_N)$ ($S$ and $N$ refer to SC and normal states, respectively), which can be expressed as a momentum weighted integral of the dynamical structure factor, which is proportional to neutron intensity:
\begin{equation}
\langle \mathbf{S}_i\cdot\mathbf{S}_j\rangle=3J\int\frac{d\omega}\pi\int\frac{d2q}{(2\pi)^2}S(\mathbf{q},\omega)(\cos(q_x)+cos(q_y)).
\end{equation}


The momentum weight factor is optimal at $\mathbf{q}=(\pi,\pi)$, which implies that shifting spectral weight from incommensurate positions towards the commensurate position lowers the magnetic exchange energy. As sketched in figure \ref{fig:sketch}, opening a spin-gap would increase the exchange energy  in a system with excitations dispersing away from the commensurate point, leave the exchange energy unchanged in a system with energy independent incommensurate fluctuations (such as our Fe$_{1.05}$Te$_{0.7}$Se$_{0.3}$ sample), but would lower the exchange energy for systems showing inwards dispersion from incommensurate wave-vectors towards the commensurate point (such as our Fe$_{1.02}$Te$_{0.7}$Se$_{0.3}$ sample).

The recent observation of an hour-glass dispersion with $E_{hg}=14$~meV in insulating, non-SC, La$_{5/3}$Sr$_{1/3}$CoO$_4$ \cite{boothroyd2011} manifest that it is not a sufficient condition for high-temperature superconductivity. Rather it places an upper limit on the spin-gap and henceforth on the SC transition temperature, since opening a spin gap larger than $E_{hg}$ would shift spectral weight back away from $(\pi,\pi)$ .  Both copper-oxide and Fe-based systems - including our sample - obey $E_{gap}\simeq4k_BT_c$ (except for highly under-doped cuprates, where $E_{gap}$ is further reduced due to softening of the $d$-wave SC gap around the nodal point \cite{chang2007}). A similar attempt of scaling $E_{hg}\simeq5.3k_BT_c$ holds for YBCO and  Fe$_{1.02}$Te$_{0.7}$Se$_{0.3}$, where the spin gap is squeezed up towards $E_{hg}$, but breaks down \emph{e.g.} for the La$_{2-x}$Sr$_x$CuO$_4$ family where other effects must suppress $T_c$ and hence $E_{gap}$.

We therefore conclude that existence of an hour-glass shaped dispersion is a necessary condition for high-temperature superconductivity of the nature found in iron- and copper-oxide based materials. This implies that the mechanism for superconductivity is a lowering of the magnetic exchange energy through shifting of spectral weight towards the commensurate point, and we conjecture that the energy $E_{hg}$ of the commensurate point in this hour-glass dispersion imposes a maximum possible transition temperature $T_c^{max}\simeq E_{hg}/5.3k_B$.
If a new family of high-temperature superconductors is discovered, measuring $E_{hg}$ will provide an estimate for the maximum achievable $T_c$ and hence provide important guidance as to whether further compositional exploration within that family may be futile or fruitful.

\section*{References}
\bibliographystyle{prsty}
\bibliography{FeTeSe}

\section{Acknowledgements}
We acknowledge useful discussions with N. B. Christensen, B. Normand and D. J. Scalapino. This work was supported by the Swiss National Science Foundation and MaNEP. Preliminary neutron scattering measurements were done at SINQ, PSI, Switzerland.

\end{document}